\documentstyle[epsf,twocolumn,prl,aps]{revtex}

\begin{document}
\draft


\twocolumn[\hsize\textwidth\columnwidth\hsize\csname @twocolumnfalse\endcsname

\title{
Thermodynamics of the bilinear-biquadratic spin one Heisenberg chain
}

\author{Jizhong Lou, Tao Xiang and Zhaobin Su }

\address{
Institute of Theoretical Physics, Chinese Academy of Sciences, P. O. Box 2735, 
	Beijing 100080, People's Republic of China
}
\date{ \today }

\maketitle

\begin{abstract}
The magnetic susceptibility and specific heat of the one-dimensional 
$S=1$ bilinear-biquadratic Heisenberg model are calculated using the 
transfer matrix renormalization group. By comparing the results with 
the experimental data of ${\rm LiVGe_2O_6}$ measured by Millet et al. 
(Phys. Rev. Lett. {\bf 83}, 4176 (1999)), we find that
the susceptibility data of this material, 
after subtracting the impurity contribution, 
can be quantitatively explained with this model. 
The biquadratic exchange interaction in this material is found to 
be ferromagnetic, i.e. with a positive coupling constant.
\end{abstract}


] 

The quantum spin chains have been the subject of many theoretical and
experimental studied since the conjecture made by Haldane \cite{haldane83}
that the antiferromagnetic Heisenberg model has a finite excitation gap for
integer spins. The model which has been intensively used to investigate the
physics behind the Haldane's conjecture is the isotropic spin one Heisenberg
Hamiltonian with both bilinear and biquadratic spin interactions:

\begin{equation}
H=J\sum_i\left[ {\bf S}_i\cdot {\bf S}_{i+1}+\gamma ({\bf S}_i\cdot {\bf S}
_{i+1})^2\right] .  \label{model}
\end{equation}
For most of the existing quasi-one-dimensional (1D) $S=1$ materials, the
biquadratic term is very small compared with the bilinear term as well as
the uniaxial anisotropy. This model was therefore generally thought to be of
pure theoretical interest. However, recently Millet et al. \cite{millet99}
found that the magnetic susceptibility of a new quasi-1D $S=1$ system, the
vanadium oxide ${\rm LiVGe}_2{\rm O}_6$, shows a few interesting features which
are absent in other $S=1$ materials. They argued that both the interchain
coupling and the uniaxial anisotropy are too small to create these features
and suggested that the biquadratic term plays an important role in this
material.

In this paper, we present a theoretical study for the thermodynamics of the
bilinear-biquadratic spin chain (\ref{model}) with $J>0$. We have calculated
the magnetic susceptibility and specific heat of this model using the
transfer matrix renormalization group (TMRG) method \cite
{bursill96,wang97,Xiang,Shibata}. By comparing with the experimental data of 
${\rm LiVGe}_2{\rm O}_6$, we find that the measured susceptibility, after
subtracting the impurity contribution, can be quantitatively fitted by the
numerical result with $\gamma =1/6$. This shows that the spin dynamics of $
{\rm LiVGe}_2{\rm O}_6$ can indeed be described by the Hamiltonian (\ref
{model}), in agreement with Millet et al. \cite{millet99}. However, the
value of $\gamma $ needed for fitting the experimental data is different
from that suggested by Millet et al. \cite{millet99}.

Let us first consider the properties of the ground state. It is
known that when $\gamma =-1$ and $1$, the model (\ref{model}) can be solved
rigorously by the Bethe Ansatz \cite{BetheAnsatz1,BetheAnsatz2}. Between
these two soluble points, the system is in the Haldane phase. In this phase,
the ground state is a non-magnetic singlet with a finite energy gap in
excitations. In particular, when $\gamma $ is between $-1$ and $\gamma
_{ic}\approx 0.41$, the low energy physics of this model can be understood
from the valence bond solid (VBS) model proposed by Affleck {\it et al. }
\cite{affleck87}. In this model, each site on the chain is occupied by two $
S=1/2$ spins and the ground state is formed by the bonding of two $S=1/2$
spins from adjacent sites. These singlet bonds must be broken in order to
excite the system and this leads to a non-zero energy gap in the low-lying
spectrum. This picture has been confirmed experimentally \cite
{Hagiawara90,Glarum91} as well as numerically \cite{dmrg1}. At $\gamma _{ic}$
, the ground state undergoes a commensurate-incommensurate transition and
the critical exponent for the magnetization changes from $1/2$ below $\gamma
_{ic}$ to $1/4$ above $\gamma _{ic}$\cite{Golinelli,Okunishi}. Between $
\gamma _{ic}$ and $1$, the system is in the incommensurate phase and the
incommensurate peak in the spin form factor $S(q)$ of the ground state moves
continuously from $\pi $ to $2\pi /3$ as $\gamma $ increases from $\gamma
_{ic}$ to $1$ \cite{xiang93,bursill95}. Above $\gamma =1$, the true nature of 
the ground state is still controversial \cite{xiang93,bursill95,schmitt98}. 
Some works \cite{xiang93,bursill95} suggest that it might be in a trimerized phase. 
When $\gamma <-1$, the ground state is doubly degenerate and dimerized.

The TMRG is a finite temperature extension of the powerful density matrix
renormalization group method \cite{White92}. A detailed introduction to this
method can be found in references \cite{bursill96,wang97,Xiang,Shibata}. The
TMRG method handles directly infinite spin chains and thus there is no
finite system size effects. To calculate the spin susceptibility, 
we first evaluate
the magnetization $M$ of the system with a small external field $B$, and
then from the ratio $M/B$ we obtain the value of the susceptibility. The
specific heat is evaluated from the numerical derivative of the internal
energy with respect to temperature. At low temperatures, since the specific
heat is very small, the relative error of the specific heat may become quite
large. In most of our calculations 100 states are retained.

\begin{figure}[ht]
 \epsfxsize=3.0 in\centerline{\epsffile{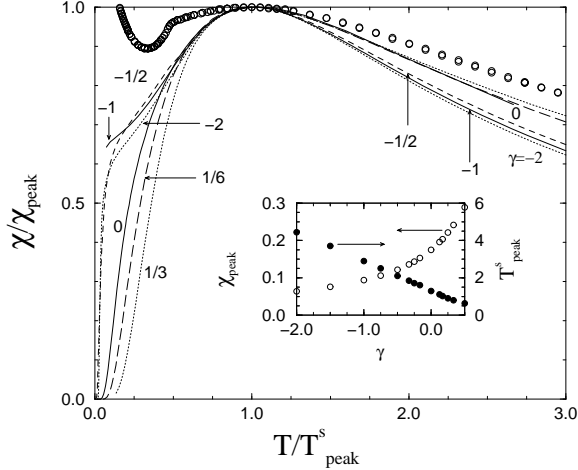}}
\caption[]{
The normalized spin susceptibility $\chi/\chi_{\rm peak}$ 
as a function of $T/T_{\rm peak}$ at zero field.
The experimental data of ${\rm LiVGe_2O_6}$ obtained by
Millet et al  \onlinecite{millet99} (empty circles) 
are also shown for comparison. The inset shows the 
$\gamma$ dependence of the peak susceptibility $\chi_{peak}$ 
(empty circles)
and the peak temperature $T_{\rm peak}^s$ (filled circles). 
$J$ is set to unit. 
}
\label{fig1}
\end{figure}

Figure \ref{fig1} shows the zero-field 
spin susceptibility $\chi (T)$ normalized by the
its peak value $\chi _{{\rm peak}}$ as a function of the normalized
temperature $T/T_{{\rm peak}}^s$ for a set of $\gamma $, where $T_{{\rm peak}
}^s$ is the temperature of $\chi _{{\rm peak}}$. Above $T_{{\rm peak}}^s$, $
\chi (T)/\chi _{{\rm peak}}$ behaves similarly for all the curves shown in
the figure. When $\gamma $ is positive, $\chi (T)$ drops quickly below $T_{
{\rm peak}}^s$. This is because the energy gap in this parameter regime is
very large. As $\gamma $ becomes negative, $\chi (T)$ just below $T_{{\rm 
peak}}^s$ tends to become flatter. At $\gamma =-1$, there is no gap in the
excitation spectrum, $\chi (T)$ shows a small positive curvature at low
temperatures, as in the $S=1/2$ Heisenberg chain. 

The inset of Figure \ref{fig1} shows the $\gamma $ dependence of $\chi _{
{\rm peak}}$ and $T_{{\rm peak}}^s$. The increase of $\chi _{{\rm peak}}$
with $\gamma $ indicates that the susceptibility becomes larger when $\gamma 
$ moves from the dimerized phase to the Haldane phase. This is consistent
with the picture that in the dimerized phase the spin is frozen by forming
rather rigid spin singlet, while in the Haldane phase the spin is relatively
free above the Haldane gap. The peak temperature $T_{{\rm peak}}^s$ drops
almost linearly with $\gamma $. The slope of this drop is about $1.6J$ per
unit $\gamma $.

\begin{figure}[ht]
\epsfxsize=2.7 in\centerline{\epsffile{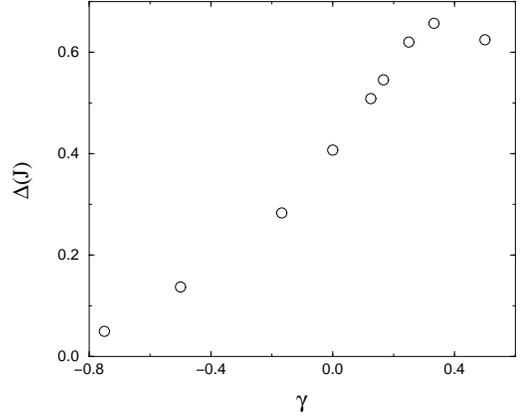}}
\caption[]{
The energy gap as a function of $\gamma$. 
}
\label{fig2}
\end{figure}

In a gaped phase, the low-lying excitation has approximately the energy
dispersion 
\begin{equation}
\varepsilon _k=\Delta +\frac{v^2}{2\Delta }\left( k-k_0\right) ^2+O\left(
\left( k-k_0\right) ^3\right) ,  \label{dis}
\end{equation}
where $k_0$ is the wavevector of the excitation minimum, $\Delta $ is the
energy gap and $v$ the spin velocity. When $T\ll \Delta $, it can be shown
that $\chi (T)$ has the form\cite{xiang98}
\begin{equation}
\chi (T)\approx \lambda \sqrt{\frac \Delta T}e^{-\Delta /T},  \label{pure}
\end{equation}
where $\lambda $ is a $T$-independent parameter. From the fit of the low
temperature TMRG results of $\chi (T)$ with this equation, we can estimate
the value of $\Delta $. The result of $\Delta $ we obtained is shown in
Figure \ref{fig2}. The maximum energy gap is $\sim 2J/3$, located at $\gamma
=$ $1/3$. Our results agree with other numerical 
studies\cite{schmitt98,scholl96}. 

\begin{figure}[ht]
\epsfxsize=2.4 in\centerline{\epsffile{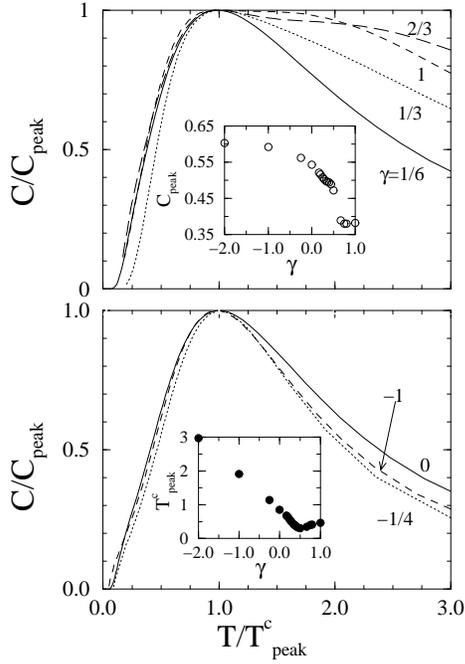}}
\caption[]{
The normalized specific heat $C/C_{\rm peak}$ as a function of $T/T_{\rm peak}^c$. 
The upper and lower panels are for $\gamma$ larger and smaller than zero, respectively. 
The upper inset shows the peak specific heat $C_{\rm peak}$ and the lower one 
shows  the corresponding 
temperature $T_{\rm peak}^c$. $J$ is set to unit.  
}
\label{fig3}
\end{figure}

Figure \ref{fig3} shows the temperature dependence of the
specific heat $C(T)$ for a set of $\gamma $. The inset of the figure shows
the $\gamma $ dependence of the peak value of the specific heat, $C_{{\rm 
peak}}$, and the peak temperature $T_{{\rm peak}}^c$. Compared with $T_{{\rm 
peak}}^s$, $T_{{\rm peak}}^c$ behaves quite differently. It drops with
increasing $\gamma $ when $\gamma <1/2$ and becomes almost a constant when $
\gamma >1/2$. Below the peak temperature, $C/C_{{\rm peak}}$ shows quite
similar behavior for all the curves shown in the figure except at very low
temperatures. Since there is no energy gap at $\gamma =\pm 1$, $C(T)$ at
these two points approaches to zero linearly with decreasing $T$. However,
for other cases, $C(T)$ decays exponentially at low temperatures. 
For the two exact solvable point $\gamma= \pm 1$, exact results 
are available\cite{sacramento}, the specific heat 
vanishes linearly at low temperature. However, due to large errors at
low temperatures, our results do not show this behavior clearly.
Above the peak temperature, $C/C_{{\rm peak}}$ drops quickly for 
negative $\gamma $.
However, when $\gamma $ becomes bigger, in particular in the incommensurate
phase ($\gamma =2/3$ and $1$), $C(T)$ shows a weak and broadened peak above $
T_{{\rm peak}}^c$. It seems that there is a new excitation mode accumulated
at low energies in the incommensurate state.

Now let us compare the numerical results with the spin susceptibility data $
\chi _{\exp }$ of ${\rm LiVGe}_2{\rm O}_6$ measured by Millet et al. on a
powder sample \cite{millet99}. As mentioned in \cite{millet99}, two
extraordinary features appear in $\chi _{\exp }$. One is the slow drop of $
\chi _{\exp }$ on both sides of the susceptibility peak, and the other is
the abrupt drop of $\chi _{\exp }$ below $22K$ with a sharp upturn below $15K
$. The first feature, in particular the slow drop of $\chi _{\exp }$ below
the peak temperature, is reminiscent of a gapless system. The second feature
of $\chi _{\exp }$ is typical of a spin-Peierls system with impurities, such
as in $Zn$ doped ${\rm CuGeO}_3$ \cite{Hase93}. These features have led
Millet et al. to interpret ${\rm LiVGe}_2{\rm O}_6$ as a nearly gapless $S=1$
spin chain with the spin-Peierls instability. However, 
whether the abrupt drop of $\chi _{\exp }$ at $22K$ is really due 
to a spin-Peierls transition is
still an open question. 

The sharp upturn of $\chi _{\exp }$ at low temperatures indicates that the
impurity contribution is strong. To see how strong the impurity effect is,
let us first do a comparison without subtracting the impurity contribution
in $\chi _{\exp }$. In Figure 1, the measured susceptibility $\chi _{\exp }$
normalized by its peak value at about $47K$ is compared with the TMRG
results discussed previously. The disagreement between the theoretical and
experimental results indicates that the impurity effect is too strong to be
ignored even at high temperatures. 

The susceptibility of dilute magnetic impurities generally has a Curie-Weiss
form 
\begin{equation}
\chi _{{\rm imp}}=\frac{C^{\prime }}{T+\theta ^{\prime }+\alpha T^{-1}},
\end{equation}
where $C^{\prime }$ is proportional to the impurity concentration and the
square of the effective $g$-factor of the impurity and $\theta ^{\prime }$
is a measure for the interaction among impurities. The $\alpha T^{-1}$ term
in $\chi _{{\rm imp}}$ is the leading order correction to the Curie-Wess
term $C^{\prime }/\left( T+\theta ^{\prime }\right) $ due to the finite
magnetic field. If there is no interaction between impurities, $\alpha
=\left( 2S^{\prime 2}+2S^{\prime }+1\right) \left( g^{\prime }\mu
_BB/k_B\right) ^2/10$ with $g^{\prime }$ and $S^{\prime }$ the effective
g-factor and spin of impurities. This term is not important at high
temperatures. But when the temperature becomes comparable with the level
splitting of an impurity spin in a magnetic field, this term becomes
important. It prevents $\chi _{{\rm imp}}$ from being divergent at low
temperatures. $\alpha $ is typically of order $1K^2$ when $B=1T$.

At very low temperatures the measured susceptibility is a sum of $\chi _{
{\rm imp}}$ and $\chi (T)$ given by Eq. (\ref{pure}), i.e. 
\begin{equation}
\chi _{\exp }(T)=\chi _{{\rm imp}}+\lambda \sqrt{\frac \Delta T}e^{-\Delta
/T}.  \label{chi0}
\end{equation}
By fitting the low temperature experimental data below $15K$ with this
equation, we find that $C^{\prime }=0.115cm^3K/mol$, $\theta ^{\prime
}=14.1K $, $\alpha =2.18K^2$, $\lambda =0.0063cm^3/mol$ and $\Delta =36K$.
These parameters show that not only the contribution from impurities to $
\chi _{\exp }$ is large as expected, but also the interaction among
impurities is strong at low temperatures. There is no simple explanation for
such a strong correlation among impurities. 
Clearly this is an important problem which should be further
investigated both theoretically and experimentally.

\begin{figure}[ht]
\epsfxsize=3.0 in\centerline{\epsffile{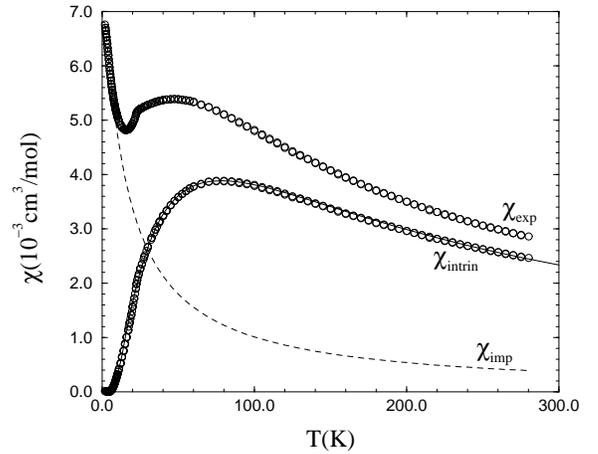}}
\caption[]{
Comparison of the TMRG result (solid line) of the spin susceptibility 
of the $S=1$ bilinear-biquadratic model 
with $J=73K$ and $\gamma =1/6$ with the experimental data of 
${\rm LiVGe_2O_6}$ \onlinecite{millet99}. 
$\chi_{\rm intrin}$ is the experimental data after 
subtracting the impurity contribution 
$\chi_{\rm imp}$ from $\chi_{\rm exp}$ . 
}
\label{fig4}
\end{figure}

By subtracting the impurity contribution from $\chi _{\exp }$, we obtain the
intrinsic susceptibility $\chi _{{\rm intrin}}$ of ${\rm LiVGe}_2{\rm O}_6$.
The result of $\chi _{{\rm intrin}}$ together with the raw data $\chi _{\exp
}$ and $\chi _{{\rm imp}}$ is shown in Figure \ref{fig4}. After the
subtraction, the abrupt drop of $\chi _{\exp }$ at $22K$ becomes less
distinct, but the change in the slope is still visible. The most significant
change of $\chi _{{\rm intrin}}$ compared with $\chi _{\exp }$ is that the
peak shifts to a higher temperature and the drop below the peak temperature
becomes more rapidly. By comparing in detail the normalized $\chi _{{\rm 
intrin}}$ with the theoretical results, we find that $\chi _{{\rm intrin}}$
can be well fitted by the numerical curve with $\gamma =1/6$ (Figure \ref
{fig4}). This shows that the biquadratic term in model (\ref{model}) does
have an important contribution to the low energy spin dynamics of ${\rm LiVGe
}_2{\rm O}_6$, in agreement with Millet et al. \cite{millet99}. However, the
value of $\gamma$ which gives the best fit, in particular its sign, is
different from that suggested in Ref. \cite{millet99}. A detailed 
comparison indicates that $\chi _{{\rm intrin}}$ lies between the 
theoretical curves for $\gamma=1/4$ and  $1/8$ in the whole temperature 
region. Thus the uncertainty in the value of $\gamma_c$ is very small. 
The result at $\gamma_c \sim -1$ suggested in Ref. \onlinecite{millet99} 
does not fit the experiment data.

At $\gamma =1/6$, the peak temperature is $T_{{\rm peak}}^s=1.025J$. Setting
this $T_{{\rm peak}}^s$ equal to the peak temperature of $\chi _{{\rm intrin}
}$, we find that $J\sim 73K$. Compared with the gap value $\Delta =36K$
obtained previously, we have $\Delta \sim 0.49J$. This value of $\Delta $ is
rather close to the Haldane gap, $0.54J$, of the Hamiltonian (\ref{model})
at $\gamma =1/6$ (Figure \ref{fig2}). This suggests that the low energy 
spin excitations are gapped and the change of the slope at 22K 
is not due to a spin-Peierls transition.

We have also compared $\chi _{{\rm intrin}}$ with the spin susceptibility of
the $S=1$ Heisenberg model with uniaxial single-ion anisotropy but without
the biquadratic term\cite{Coombes98}, namely the model $H=J\sum_i\left[ {\bf 
S}_i\cdot {\bf S}_{i+1}+D\sum_iS_{iz}^2\right] $. However, in the parameter
region which might be physically relevant, $-1/2<D<1/2$, we find that none
of the numerical curves fits $\chi _{{\rm intrin}}$ in the whole temperature
range. This shows that the uniaxial anisotropy in ${\rm LiVGe}_2{\rm O}_6$
is indeed very small, in agreement with the analysis of Millet et al. \cite
{millet99}.

The above analysis confirms the importance of the biquadratic
exchange interaction in ${\rm LiVGe}_2{\rm O}_6$. On the other hand, it also
raises some new questions. In the argument given by Millet et al., the
biquadratic term comes from in fourth order since at 
second order the antiferromagnetic and ferromagnetic terms 
partially cancel. However, the coefficient of this biquadratic term is
negative (i.e. $\gamma <0$) according to their calculation, in contrast with
the result we obtain. To resolve this disagreement, further investigation
into the electronic structure of ${\rm LiVGe}_2{\rm O}_6$ is needed. More
detailed measurements with high quality single crystals would also help 
clarify the impurity effect as well as the nature of the anomaly
at $22K$ in this material. In a $S=1$ Heisenberg chain, the
localized non-magnetic impurity may induce mid-gap states within the Haldane
gap \cite{Ditusa94,Soensen,wangxq,Lou}. A better understanding of the physical
properties of these mid-gap states would also be helpful for further understanding
the thermodynamics of ${\rm LiVGe}_2{\rm O}_6$ at low temperatures.

In summary, the thermodynamic properties of the bilinear and biquadratic
Heisenberg model have been studied and compared with the experiments. The
measured susceptibility data of ${\rm LiVGe}_2{\rm O}_6$, after subtracting
the impurity contribution, can be quantitatively explained by the model (\ref
{model}) with $\gamma =1/6$.

We wish to thank F. Mila and F. C. Zhang for sending us the experimental
data, and M. W. Long, N. d'Ambrumenil and G. A. Gehring for useful discussions. TX
acknowledges the hospitality of the Isaac Newton Institute of the University
of Cambridge, where this work was completed. This work was supported in part
by the National Natural Science Foundation of China and by the Special Funds
for Major State Basic Research Projects of China .


\begin{references}
\bibitem{haldane83}  F. D. M. Haldane, Phys. Rev. Lett. {\bf 50}, 1153
(1983); Phys. Lett. {\bf 93A}, 464 (1983).

\bibitem{millet99}  P. Millet, F. Mila, F. C. Zhang, M. Mambrini, A. B. Van
Oosten, V. A. Pashchenko, A. Sulpice, and A. Stepanov, Phys. Rev. Lett. {\bf 
83}, 4176 (1999).

\bibitem{bursill96}  R. J. Bursill, T. Xiang and G. A. Gehring, J. Phys.
Cond. Matt. {\bf 8}, L583 (1996).

\bibitem{wang97}  X. Wang and T. Xiang, Phys. Rev. B {\bf 56}, 5061 (1997).

\bibitem{Xiang}  T. Xiang and X. Wang, ``Finite temperature and momentum
space DMRG'' in {\it Lecture Note in Physics:} Density-Matrix
Renormalization, Edited by I. Peschel {\it et al., } Springer (1999).

\bibitem{Shibata}  N. Shibata, J. Phys. Soc. Jpn. {\bf 66}, 2221 (1997).

\bibitem{BetheAnsatz1}  L. A. Takhtajan, Phys. Lett. {\bf 87A}, 479 (1982);
H. M. Babudjian, Nucl. Phys. B{\bf 215}, 317 (1983).

\bibitem{BetheAnsatz2}  G. V. Uimin, JETP Lett. {\bf 12}, 225 (1970); C. K.
Lai, J. Math. Phys. {\bf 15}, 1675 (1974); B. Sutherland, Phys. Rev. B {\bf 
12}, 3795 (1975).

\bibitem{affleck87}  I. Affleck, T. Kennedy, E. H. Lieb and H. Tasaki, Phys.
Rev. Lett. {\bf 59}, 799 (1987); Commun. Math. Phys. {\bf 115}, 477 (1988).

\bibitem{Hagiawara90}  I. Hagiawara, K. Katsumata, I. Affleck, B.I. Halperin
and J.P. Renard, Phys. Rev. Lett. {\bf 65}, 3181 (1990).

\bibitem{Glarum91}  S. H. Glarum, S. Geschwind, K. M. Lee, M. L. Kaplan, and
J. Michel, Phys. Rev. Lett. {\bf 67}, 1614 (1991).

\bibitem{dmrg1}  S. R.\ White and D. A.\ Huse, Phys. Rev. B {\bf 48} 3844
(1993).

\bibitem{Golinelli}  O. Golinelli, Th. Jolic\oe ur and E. S. S\o rensen,
Euro. Phys. B {\bf 11}, 199 (1999).

\bibitem{Okunishi}  K. Okunishi, Y. Hieida, and Y. Akutsu, Phys. Rev. B {\bf 
59}, 6068 (1999).

\bibitem{xiang93}  T. Xiang and G. A. Gehring, Phys. Rev. B {\bf 48}, 303
(1993).

\bibitem{bursill95}  R. J. Bursill, T. Xiang and G. A. Gehring, J. Phys. A 
{\bf 28}, 2109 (1995).

\bibitem{schmitt98} A. Schmitt, K.-H. M\"utter, M. Karbach, Y. Yu, and G. M\"uller,
Phys. Rev. B {\bf 58}, 5498 (1998).

\bibitem{White92}  S. R.\ White, Phys. Rev. Lett. {\bf 69}, 2863 (1992).

\bibitem{xiang98} T. Xiang, Phys. Rev. B {\bf 58}, 9142 (1998) 
and references therein. 

\bibitem{scholl96} U. Schollw\"ock, T. Jolic{\oe}ur, and T. Garel,
Phys. Rev. B {\bf 53}, 3304 (1996). 

\bibitem{sacramento} Z. Sacramento, Phys. B {\bf 94}, 347 (1994).

\bibitem{Hase93}  M. Hase, I. Terasaki, Y. Sasago, K. Uchinokura, H. Obara,
Phys. Rev. Lett. {\bf 71}, 4059 (1993).

\bibitem{Coombes98}  D. Coombes, T. Xiang, and G. A. Gehring, J. Phys.
Condens. Matter {\bf 10}, L159 (1998).

\bibitem{Ditusa94}  J. F. Ditusa, S. W. Cheong, J. H. Park, G. Aeppli, C.
Broholm and C. T. Chen, Phys. Rev. Lett. {\bf 73}, 1857(1994).

\bibitem{Soensen}  E. S.\ S\o rensen and I. Affleck, Phys. Rev. B {\bf 51},
16115 (1995).

\bibitem{wangxq} X. Wang and S. Mallwitz,  
                 Phys. Rev. B {\bf 53}, R492 (1996).

\bibitem{Lou}  W. Wang, S. Qin, Z. Lu, Z. B. Su, and L. Yu, Phys. Rev. B 
{\bf 53}, 40 (1996); J. Lou, S. Qin, Z. B. Su and L. Yu, {\it ibid }{\bf 58}
, 12672 (1998).
\end{references}
\end{document}